# Upper bound on the disordered density of sphere packing and the Kepler Conjecture

Jozsef Garai

jozsef.garai@fiu.edu

**Abstract.**

The average distance of the equal hard spheres is introduced to evaluate the density of a given arrangement. The absolute smallest value is two radii because the spheres can not be closer to each other than their diameter. I call the density relating to the two radii distance to absolute highest density. The absolute densest arrangement of two, three and four spheres is defined, which gives the absolute highest density in one, two and three dimensions. The absolute highest density of equal spheres in three dimensions is the tetrahedron formed by the centers of four spheres touching each other. The density of the enclosed tetrahedron is 0.7796, which is the absolute highest density of equal spheres in three dimensions. The density of this tetrahedron unit can be maintained only locally because the tetrahedron units can not be expanded to form a tightly packed arrangement in $\mathbb{R}^3$. The maximum number of tetrahedron units that one sphere is able to accommodate is twenty corresponding to the density of 0.684. The only compatible formation of equal spheres which can be mixed with tetrahedron is octahedron. Certain mixture of these two units might result in higher density than 0.684. In order to mix the tetrahedron and octahedron units certain geometrical constrains must be satisfied. It is shown that the only possible mixture of tetrahedrons and octahedrons units is the one which accommodates eight tetrahedron and six octahedron vertexes. This arrangement corresponds to FCC. By demonstrating that no highest density than FCC can be formed an alternative proof for the Kepler conjecture is provided. It is suggested that there is a density gap between the FCC density and the highest density of disordered arrangements because there is no other tetrahedron and octahedron configuration exists between the FCC and the density of the icosahedrons configuration. It is also suggested that the icosahedrons configuration with its 0.684 density represents the upper bound for disordered arrangements.

# I. INTRODUCTION

Finding an answer to the question, "What would be the densest arrangement of equal spheres in three dimensional spaces?" has been challenging mathematicians for centuries. Johannes Kepler stated in 1611 that the face centered cubic arrangement (FCC) [1] gives the densest possible packing, however he never proved his statement [2]. It has been proven by Gauss that the densest arrangement for lattice packing in three dimensions is FCC [3].

Several upper bounds on the density of spheres

0.884 Blichfeld [4]

0.835 Blichfeld [5]

0.828 Rankin [6]

0.7796 Rogers [7, 8]

0.77844 Lindsley [9]

0.77836 Muder [10]

0.7731 Muder [11] have been proposed [12]. It is interesting to note that there is a decreasing trend in the upper bound value.

In 1990, and again in his revised version in 1993, Wu_Yi Hsiang [13] is clamed that he solved the Kepler's Conjecture. Experts in the field of sphere packing found errors in his solution and determine that many details had been omitted [14, 15]. In 1997 Thomas C. Hales announced that he had a proof of the Kepler conjecture [16-19]. The overall correctness of this work is widely accepted; however, the proof with its long computer calculations is still a matter of discussion [20-23] and waits for universal acceptance.

The density of disordered spheres is studied by computer simulations [24-34] and physical experiments [15, 35-39]. Both types of investigations give the most probable density for disordered spheres around 0.64 which is called random closed packing limit. The difference between the most probably density of disordered packing and FCC is about 13.5% indicating that a gap between FCC and disordered packing might exist. In this study the packing problem of equal spheres is revisited. Explanation for the most probable density of disordered spheres is proposed. Simple proof for the Kepler Conjecture and for the upper bound on the disordered density of sphere packing is presented.

# II. ABSOLUTE CLOSEST PACKING IN ONE, TWO AND THREE DIMENSION

**Postulate 1**

*The highest density of two spheres is reached when the surface of the spheres touches each other.*

Let $r_1$ be the position vector pointing to the center of sphere 1, $r_2$ the position vector of the center of sphere 2 and R be the radius of the spheres. Postulate 1 can be stated then as:

$$\left\|\overrightarrow{r_1 r_2}\right\| = \left\|\overrightarrow{r_2 r_1}\right\| \equiv 2R \tag{1}$$

If the spheres are not touching each other then the distance between the two centers of the spheres is bigger than 2R and can be calculated as:

$$d = \left\|\overrightarrow{r_1 r_2}\right\| = \sqrt{(x_2 - x_1)^2 (y_2 - y_1)^2 (z_2 - z_1)^2} \;. \tag{2}$$

For more than two spheres the average distance $\left[\overline{d}\right]$ between the centers of the spheres can be calculated as:

$$\overline{d} = \frac{\sum_{k=1}^{n}\sum_{i=1}^{n}\left\|r_k - r_i\right\|}{n(n-1)} \;. \tag{3}$$

where n is the number of spheres. The density of the same number of spheres can be evaluated by comparing the average distance of the spheres. The lower d value corresponds to higher density. The lowest bound on the average distance is 2R when each and every sphere is in its closest packing arrangement to all of the rest of the spheres. The lowest bound $\left[\overline{d} = 2R\right]$ corresponds to the densest arrangement of the spheres:

$$\frac{\sum_{k=1}^{n}\sum_{i=1}^{n}\left\|r_{ko} - r_{io}\right\|}{n(n-1)} \equiv 2R \;\; \Rightarrow \;\; \phi^{max} \tag{4}$$

and I will call the related density to absolute highest density $\left[\phi^{max}\right]$. The absolute highest density of two spheres reiterates postulate 1. The absolute highest density of three spheres is the arrangement when the center of the spheres forms an equilateral triangle where the sides are equal with the diameter of the spheres. The absolute highest density of four spheres is the arrangement when the centers of the spheres forms a tetrahedron with sides equal with the diameter of the spheres. These absolute highest density arrangements of two, three and four spheres can be consider as the absolute highest density units in one, two and three dimensions respectively. The defined absolute closest packing units in $\mathbb{R}^1$, $\mathbb{R}^2$ and $\mathbb{R}^3$ are shown in FIG. 1. If these closest packing units are expandable infinitely then they will reproduce their absolute highest packing density which sets an upper bound on all possible density.

The units are expandable if the available space is completely occupied. The criterion of the complete occupation of space is that the volume of the sphere divided by the volume cut out by one unit from the sphere is integer.

In the one dimensional extension the unit cuts out half of the sphere and one sphere accommodates two units. Thus the fraction is integer and the closest packing unit of $\mathbb{R}^1$ is expandable. The array, built up from the closest packing units of $\mathbb{R}^1$, has the same density as the unit itself. The density of the arrangement is the absolute highest density of equal spheres $\left[\phi_{R^1}^{max}\right]$ in one dimension $[\mathbb{R}^1]$ (FIG. 2.). The density of this arrangement is

$$\phi_{R^1}^{max} = \frac{\frac{4}{3}\pi R^3}{(2R)^3} = \frac{\pi}{6} = 0.5236 . \qquad (5)$$

In two dimensions the equilateral triangle units cut out 60 degrees or $\pi/3$ from a sphere which is $1/6$ volume fraction of the sphere and one sphere accommodates six units. Thus the $\mathbb{R}^2$ units are able to completely fill the available space and the density of the units is the highest density in $\mathbb{R}^2$ (FIG. 2). The highest density in $\mathbb{R}^2$ is then

$$\phi_{R^2}^{max} = \frac{0.5\frac{4}{3}\pi R^3}{\frac{2R\sqrt{3}R}{2}2R} = \frac{\pi}{3\sqrt{3}} = 0.6046 \qquad (6)$$

The absolute closest packing arrangement reproduces the well-known hexagonal closest packing arrangement. It has been proven in two different ways [40, 41] that hexagonal closest packing is the closest packing for equal spheres in two dimensions.

The volume, cut out by a tetrahedron unit from unit a sphere $[V_{TH-sph}]$ is calculated. Integrating the volume of a unit ball $\theta$ between $\theta = 0 - 0.392\pi$ encloses two tetrahedron and one octahedron units (FIG. 3).

$$2V_{TH-sph} + V_{OH-sph} = \int_0^{0.392\pi}\int_0^{\pi}\int_0^{a} \rho^2 \sin\phi d\rho d\phi d\theta = 0.82064$$

$$\text{where} \quad 0.392\pi = 2\text{ArcSin}\left(\frac{1}{\sqrt{3}}\right) \quad \text{and} \, a = 1 \qquad (7)$$

Integrating the volume between $\theta = 0 - 0.608\pi$ cuts out the volumes of two tetrahedrons and two octahedrons (FIG. 4).

$$2V_{TH-sph} + 2V_{OH-sph} = \int_0^{0.608\pi}\int_0^{\pi}\int_0^{a} \rho^2 \sin\phi d\rho d\phi d\theta = 1.273755$$

$$\text{where} \quad 0.608\pi = 2\text{ArcSin}\left(\sqrt{\frac{2}{3}}\right) \qquad (8)$$

The integration of the entire sphere $\theta = 0 - 2\pi$ contains 8 tetrahedrons and 6 octahedrons.

$$8V_{TH-sph} + 6V_{OH-sph} = \int_0^{2\pi}\int_0^{\pi}\int_0^{a} \rho^2 \sin\phi d\rho d\phi d\theta = 4.18879 . \qquad (9)$$

The volumes cut out by the tetrahedrons and octahedrons $V_{OH-sph}$ can be calculated from any of the two Eqs. (7)-(9). Solving the equations gives the volume cut out by the tetrahedron and octahedron units from a unit ball

$$V_{TH-sph} \cong 0.183762 \quad \text{and} \quad V_{OH-sph} \cong 0.453116. \quad (10)$$

The number of tetrahedron units that one sphere can accommodate is then

$$n_{TH-sph} = \frac{V_{sph}}{V_{TH-sph}} \cong \frac{4.18879}{0.183762} \cong 22.79. \quad (11)$$

The number of tetrahedrons accommodated by a sphere $[n_{TH-sph}]$ is not an integer; therefore, the tetrahedron units can not fill the available space tightly.

The volumes of tetrahedron $[V_{TH}]$ and octahedron $[V_{OH}]$ units formed by the centers of the spheres are

$$V_{TH} = \frac{\sqrt{8}}{3} = 0.942809 \quad \text{and} \quad V_{OH} = \frac{8\sqrt{2}}{3} = 3.771236. \quad (12)$$

The densities of the tetrahedron and octahedron units are

$$\phi_{TH} = \frac{4 \times V_{TH-sph}}{V_{TH}} = 0.7796 \quad \text{and} \quad \phi_{OH} = \frac{6 \times V_{OH-sph}}{V_{OH}} = 0.7209 \quad (13)$$

The density of the tetrahedron enclosed by the centers of four spheres corresponds to the absolute highest density; therefore, the highest density possible achievable in $\mathbb{R}^3$ is

$$\phi_{\mathbb{R}^3}^{max} = \phi_{TH} = 0.7796. \quad (14)$$

This upper bound is the same as Rogers's limit [7, 8]. This absolute highest density of the $\mathbb{R}^3$ unit can not be maintained infinitely because the tetrahedron units do not form a tightly packed arrangement [Eq. (11)] (FIG. 2).

### III. CLOSEST PACKING OF DISORDERED ARRANGEMENTS

**Lemma 1**

*The highest number of tetrahedron units that one sphere is able to accommodate is 20.*

*Proof.* The tightly packed tetrahedron units with vertexes in the center of one of the sphere should form a surface which contains only equal lateral triangle faces. The number of solids containing only equilateral triangle faces is limited. Meeting 3, 4, or 5 triangles at each vertex give rise to tetrahedron, octahedron or icosahedron respectively. The radius of the circumscribed sphere (the one that touches the icosahedron at all vertices) is

$$r_{icosahedron} = \frac{a}{4}\sqrt{10 + 2\sqrt{5}} \cong 0.9510565163a \qquad (15)$$

where a is the edge length of the regular icosahedron. In order to accommodate tetrahedron units the radius should be equal to the edge length of the icosahedron. Since this condition is not satisfied it can be concluded 20 tetrahedron units can not be tightly packed. The twenty tetrahedron units can be packed in a center of a sphere but extra space between these units will remain. In case of unit spheres the edge of the tetrahedron is equal with two units (FIG. 5) and the radius of the circumscribed sphere is

$$r_{icosahedron} \cong 2 \times 0.9510565163 \cong 1.902113. \qquad (16)$$

In order to accommodate the tetrahedron units $r_{icosahedron}$ has to be increased to two. The increase of the radius to two will increase the length of the great circle by

$$\Delta l = 2\pi(2 - 1.902113) = 0.6150. \qquad (17)$$

The minimum distance required to accommodate an additional equilateral triangle with edge of two is $\sqrt{3}$. Since

$$\Delta l < \sqrt{3} \qquad (18)$$

there is no extra space to accommodate any additional tetrahedron unit. Thus the maximum number of tetrahedron units that a sphere is able to accommodate is twenty (FIG. 5) and the lemma is proved. Proving the lemma indirectly also proves the three dimensional kissing number [42-47] showing that no more than twelve spheres can touch a sphere.

Using simple geometry it can be shown that by expanding the equilateral triangle formation by $\Delta l$ [Eq. (17)] and placing a sphere into the created "hole" the sphere placed into the hole do not contribute to the density of the expanded tetrahedron formation [FIG. 5]. Assuming that each sphere accommodates 20 tetrahedron units gives the density of the icosahedron formation:

$$\phi_{rcp}^{max} = \rho_{TH} \frac{20 V_{TH-sph}}{V_{sph}} = 0.684 \qquad (19)$$

It is suggested that the twenty tetrahedron units with joint vertexes in one point represents an upper limit on the random closed packing (rcp). If there is no constrain on the position of the tetrahedrons, like the vertexes meets in one point when the density is calculated in Eq. (19), then higher density of tetrahedrons is possible [48-50].

## IV. KEPLER CONJECTURE

The only compatible arrangement of spheres which could be mixed with tetrahedrons is octahedron formation. Theoretically it might be possible that certain mixture of tetrahedron and octahedron units gives higher density than 0.684 [Eq. (19)]. The compatibility of the tetrahedron and octahedron units requires that the number of faces $[n_{face}]$ sheared by tetrahedron and octahedron units must be an integer. The number of faces shared by tetrahedrons and octahedrons is calculated as:

$$n_{face} = \frac{4n_{OH} + 3n_{TH}}{2}. \qquad (20)$$

where $n_{OH}$ is the number of the octahedron units. The individual tetrahedron and octahedron vertexes have 3 and 4 faces respectively. Thus, dividing the number of faces by 3 and 4 should also result in an integer. Applying these two phase criteria it can be shown that only the 8 tetrahedron and 6 the octahedron vertex combination is possible. The 6 octahedron and 8 tetrahedron vertexes ratio is a replica of FCC packing. Demonstrating that the pure tetrahedron unit arrangements give lower density than FCC and that the only geometrically possible mixture of tetrahedron and octahedron vertexes is the FCC arrangement proves that no higher bulk density for equal spheres is possible than FCC. Thus the Kepler conjecture is proved.

Please note that disordered packing containing only tetrahedron units can result in a locally higher density up to 0.7796. However, this highest density can not be maintained beyond the radii of two units because the tetrahedron units can not fill the available space completely; therefore, the density falls below 0.684. Between the upper bound on the random closed packing and FCC no other combination of tetrahedron and octahedron exist; therefore, it is suggested that there is a density gap between FCC and the upper bound on the random closed packing 0.7405-0.684.

## V. CONCLUSIONS

It is suggested that if the average distance of a group of spheres is 2R then the arrangement has the absolute highest density. The absolute highest densities in one, two and three dimensions are determined. The highest density in three dimensions is achieved by four spheres forming a tetrahedron. The density of this arrangement is 0.7796. The tetrahedron units can not form a tightly packed arrangement; therefore, the highest density of the tetrahedron unit can be maintained only locally.

The maximum number of tetrahedron units that one sphere is able to accommodate is twenty. The density of this arrangement is smaller than FCC. Applying the face requirements for mixing tetrahedrons and octahedrons it is shown that only the 8 tetrahedron and 6 octahedron vertex arrangement is possible. This combination reproduces the density of FCC. Increasing the number of tetrahedron units permits a locally denser arrangement but this higher density can not be maintained beyond two unit radius. The highest bulk density of equal spheres is the density of FCC thus the Kepler's conjecture is proved.

It is suggested that the accommodation of the maximum tetrahedron units is an upper bound on the disordered arrangements. It is also suggested that there is a density gap between the upper bound of the disordered arrangement and FCC because no other tetrahedron and octahedron configuration is possible then FCC.

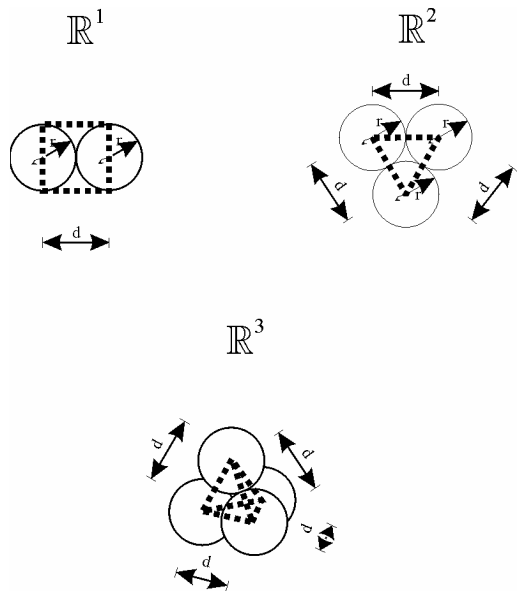

FIG. 1. The absolute densest arrangements of equal spheres in one, two and three dimensions.

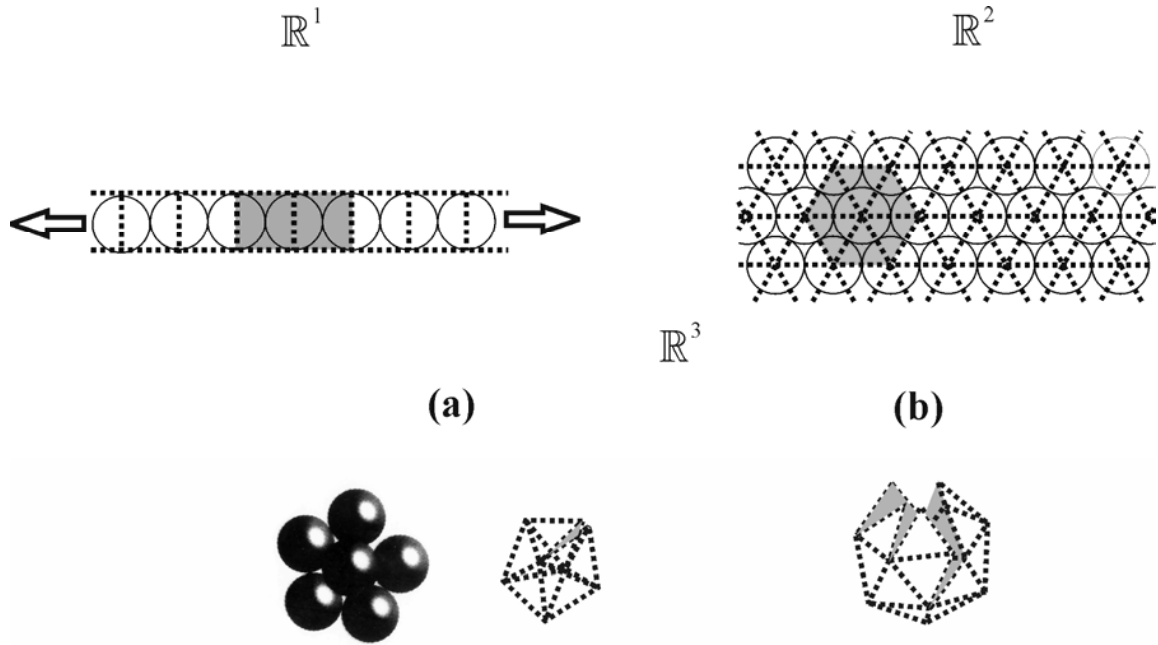

FIG. 2. The expansion of the closest packing units in three dimensions. Tetrahedron units can not form tightly packed arrangement in $\mathbb{R}^3$. (a) Arrangement of five tetrahedron units. (b) Arrangement of twenty tetrahedron units icosahedron formation.

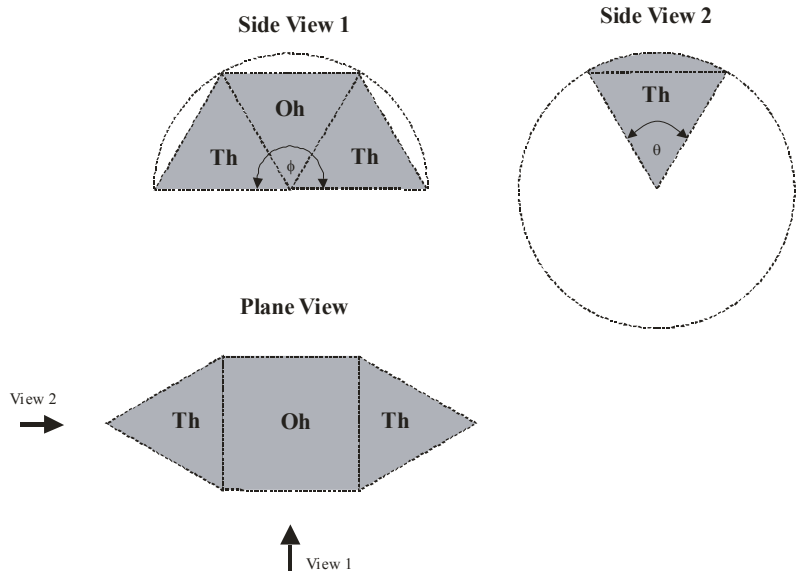

FIG. 3. Schematic arrangement of the tetrahedron and octahedron units for the integration of Equation 7.

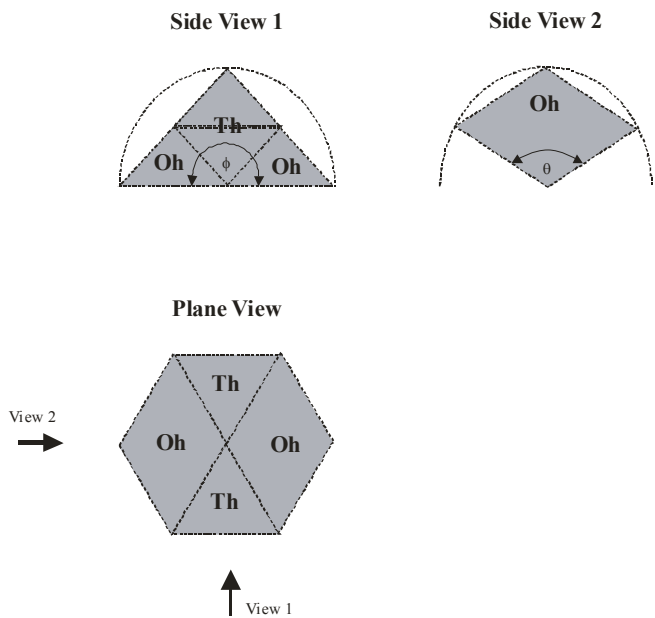

FIG. 4. Schematic arrangement of the tetrahedron and octahedron units for the integration of Equation 8.

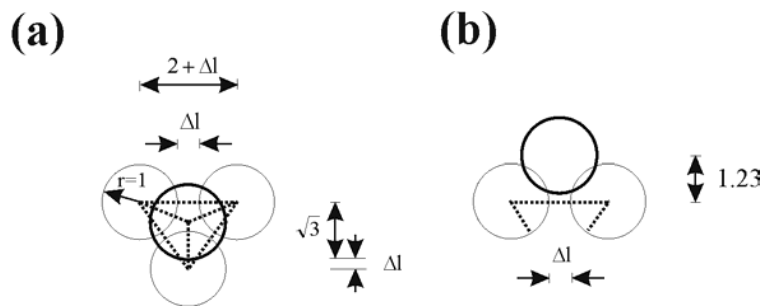

FIG. 5. Expanding the equilateral triangle formation by $\Delta l$ in two directions it can be shown that the center of the sphere placed into the hole situates outside of the tetrahedron formation; therefore, only the central sphere contributes to the density of the extra tetrahedron units. (a) Plain view of the extended tetrahedron formation. (b) Side view of the extended tetrahedron formation.